\documentclass[prl,superscriptaddress,floatfix,nofootinbib, twocolumn,aps,longbibliogrphy]{revtex4-2}

\usepackage{graphicx,amssymb,amsmath,bm,dcolumn,xcolor,mathtools}
\usepackage[percent]{overpic} % percent option lets you use percentage positions
\usepackage{tikz}
\usepackage{dblfloatfix}
\usepackage{transparent}

\usepackage[caption=false]{subfig}
\usepackage[colorlinks,linkcolor=black,citecolor=blue,urlcolor=blue]{hyperref}
\usepackage{threeparttable}
\usepackage{comment}
\usepackage{xfrac}
\usepackage{enumitem}
\usepackage[normalem]{ulem}
\usepackage{soul}
\usepackage{multirow}
\usepackage{lipsum} 
\usepackage[export]{adjustbox}

\newcommand\Xe{S}
\newcommand\Rb{K}

\newcommand\red[1]{#1}

\newcounter{parnum}
\newcommand{\parlabel}[1]{\textcolor{blue}{\textbf{\texttt{\theparnum}\addtocounter{parnum}{1}. #1}}}

\renewcommand{\parlabel}[1]{}

\begin{document}

\title{Error-correcting
transition pulses for co-located spin ensembles without frequency
selectivity}

\author{K. L. Wood}
\affiliation{Department of Physics, Arizona State University, Tempe, AZ 85281, USA}

\author{W. A. Terrano}
\email{wterrano@asu.edu}
\affiliation{Department of Physics, Arizona State University, Tempe, AZ 85281, USA}

\begin{abstract}

We present a new class of control pulses designed to transfer co-located ensembles 
without relying on frequency selectivity, thereby allowing much 
faster state-transitions. A geometric approach allows us to construct sequences which are  
robust to changes in the background 
magnetic field along multiple axes, and errors in the pulse area. \red{These pulses are extremely fast, with robustness to pulse area shown at half the quantum speed limit.}  We demonstrate these sequences on nuclear-dipole states, showing milliradian precision over several hours, 30-fold 
beyond the previous state of the art. This provides a path for extending the coherent 
integration time of ultra-long-lived nuclear-spin states to 
the fundamental limit set by their $>$10000 second lifetimes, as the limiting self-
interactions of the nuclei are suppressed in the symmetric superposition. The state-
preparation quality demonstrated here directly 
opens up 30-fold improvements in next generation tests of the standard model, 
especially tests of the symmetries of QCD and searches for dark matter; it is  
also crucial for the development of nuclear-spin based quantum memories \red{and may be 
useful in other scenarios demanding  extremely fast but robust transitions}.

\end{abstract}

\maketitle

\begin{figure}
           \includegraphics[trim={1.2cm 0.5cm 1cm 1cm}, clip, width=0.95\linewidth]{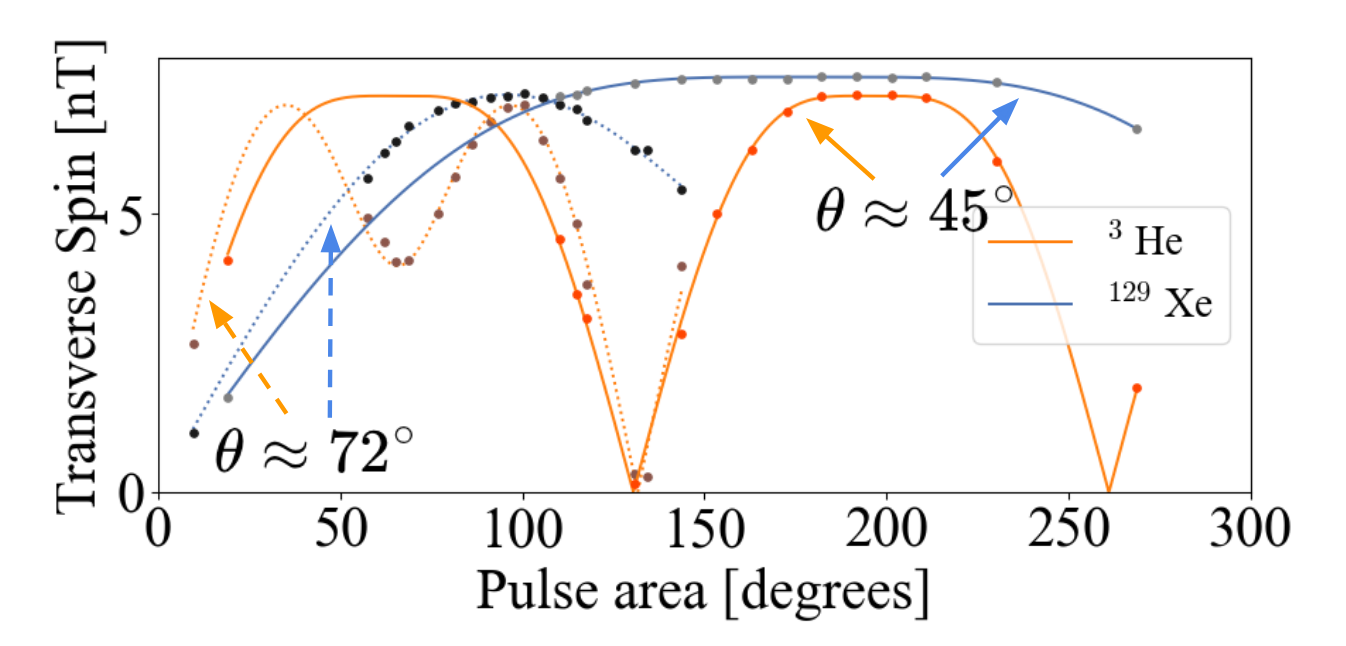}
    \caption{Joint pulses as a 
    function of pulse area. The broad flat feature shows robustness to errors in pulse 
    area for both species. Points are data, with error 
    bars from 4 measurements taken 2 
    hours apart.  The curves are single-parameter fits to 
    the state-transition model of
    equations \ref{eq: Hamiltonian} and \ref{eq: 
    Rotations}, which also define $\theta$. The polarization was the only free parameter, see equation \ref{eq: nt} for  
    conversion to nT.}
    \label{fig:45v70}
\end{figure}

Ultra-long-lived superposition states, with lifetimes well over one hundred seconds, are used 
to study fundamental physics and the symmetries of the standard model\cite{Lamoreaux1986, Lamoreaux1987, Chupp1989, 
Graner2016, Sachdeva2019, Berglund1995, Kostelecky1999, Bear2000, Brown2010, 
Gemmel2010b, Smiciklas2011, Peck2012, 
Flambaum2017, Graham2018, Wu2020, Bloch2020, Vasilakis2009, Bulatowicz2013, 
Hunter2013, 
Lee2018, Almasi2020, feng2022search} and for sensing more generally\cite{Moeller2002, Berthault2009, Mugler2010, Farooq2020}.  Applications in  quantum information processing and long-term quantum memories are being developed\cite{katz2020long, katz2022optical, katz2022quantum}.  In sensing applications, 
the metrologically relevant time-scale is 
that over which the phase evolution can be coherently integrated,
i.e. the time over which instabilities in the phase evolution are 
smaller than the uncertainties in the phase measurement. To date, the longest  ``coherent integration times'' are a few hundred seconds, while the longest $T_2$ times exceed ten thousand 
seconds\cite{Gemmel2010a}. As the coherent integration time has a 
fundamental upper limit set by the $T_2$ lifetime, there is significant room for improvement even in systems with ultra-long-lifetimes.  

Recent work has shown that higher-fidelity state preparation is the key to 
extending the integration time in the ultra-long-lived states of nuclear-spin comagnetometers\cite{Limes2019a, 
Terrano2019b}. Specifically, a non-zero longitudinal spin ${S_z = \hbar/2\,\mathrm{Tr}(\hat{\sigma}_z \rho) \equiv S 
 \sin{\alpha}}$ generates frequency shifts
which change with time as the polarization decays.  
The perfectly symmetric superposition state, i.e. with $\alpha = 0$, would be immune 
to this effect.

State-preparation fidelity of $\alpha < 10^{-3}$ is required to reach the 
fundamental limit set by the lifetime, at current sensitivities\cite{Terrano2021}. 
A variety of approaches to 
creating this state from the optically-pumped state have been tried, including multiply-resonant 
pulses\cite{Sachdeva2019}, non-adiabatic magnetic field shifts
\cite{Gemmel2010a, Sachdeva2019}, sharp pulses\cite{Limes2018a}, and transverse 
pumping\cite{Korver2015, Thrasher2019}.  These generally reached around $
\alpha\lesssim 10^{-2}$, although by and large $\alpha$ could not be measured directly.

Reliably reaching $\alpha < 10^{-3}\,$  
 will require control 
protocols that are robust against experimental 
errors over many hours. The precise problem is to design robust Hadamard gates 
(or $\pi/2$-pulses) for overlapping nuclear dipole states in low magnetic fields.  A  classic approach to making robust 
state-transitions involves composite pulses, in which $\pi$ pulses are 
inserted around various axes such that errors cancel\cite{Levitt1986, Gullion1990, Brif10a}.   
A newer approach is to dynamically stabilize the ensemble during the state-transition  
by applying a continuous stream of $\pi$-pulses\cite{Viola99a, Vitali99a, 
Souza2012b}.  Unfortunately, these strategies break 
down when working with ultra-long-lived states, motivating us to develop the  
alternative approach presented here and demonstrated on a helium-xenon comagnetometer (Fig \ref{fig:45v70}). 

The key issue is that the independent Pauli operators \{$\sigma_j^{(1)} 
\otimes \sigma_j^{(2)}$\}, $j = x,y,z$ for multiple ensembles are 
cumbersome to implement using traditional frequency selectivity when the 
transition frequencies 
are low.  In that approach, a pulse is made up of multiple Fourier 
components that independently address the overlapping ensembles, and so 
must be long enough that the Fourier components are sufficiently resolved. 
A larger problem with slow pulse-sequences is that they will violate the assumption of stationary errors, which underpins their error-correcting 
properties.  The need to control multiple ensembles with slow transitions 
is intrinsic to ultra-long-lived room-temperature states, which rely on 
weak holding fields to minimize gradient 
dephasing and the comparison of co-located ensembles to suppress background field 
drifts\cite{Chupp1988}.

 Here, we develop an alternative approach built from the limited set of  
 joint Pauli operators available when the ensembles cannot be addressed 
 independently, namely the operators {\{$
 \mathbb{I}^{(1)} \otimes \sigma_j^{(2)} + f \sigma_j^{(1)} \otimes \mathbb{I}^{(2)}$\} }, where $f$ is the ratio between the transition frequencies.  The traditional set 
 of operators allow error-correction for 
 each ensemble to be designed without regard to the other.  The faster set of operators 
 we work with here requires the joint evolution of both ensembles to be error 
 correcting. This is the first protocol capable of such joint error correction that we know of. 
 
We start with the fastest possible joint pulses and then build sequences 
that cancel errors. We present an analytic solution for joint pulses that nearly 
saturate the ``quantum speed limit'' or  time-energy 
 uncertainty principle.  Then we use bang-bang decoupling to add robustness to errors 
 in coil alignment, pulse area, and 
 background field drifts along all axes.  We validate our 
protocol on co-located ensembles of ${}^3$He and ${}^{129}$Xe spins, and demonstrate 
the 30-fold improvement in state-preparation needed for $T_2$ limited measurements 
of the nuclear dipole frequencies.  A similar improvement in next 
 generation tests of strong CP-violation and searches for dark matter is thereby made 
 possible.

Our 
control Hamiltonian consists of a sequence of   
static magnetic fields whose orientation and 
strength differs  
between each segment. Specifically  
\begin{equation}\label{eq: Hamiltonian}
    \begin{aligned}
   & H = \sum_j H_j(t); %\quad t_j \leq t < t_{j+1} 
    \quad H_j = \frac{1}{2} \vec{B}_j \cdot \left(\gamma_{1} \vec{\sigma}^{(1)} + \gamma_{2} \vec{\sigma}^{(2)} \right), 
    \end{aligned}
\end{equation}
where $\gamma_i$ and $\vec{\sigma}^{(i)}$ are the gyromagnetic ratios and Pauli operators for each ensemble. 

The state evolution can be computed as precession 
around the static field of that segment, if the duration $\tau_j$ 
is short compared to other time-scales. 
A pulse sequence is then the stack of rotations 
given by    
\begin{equation}\label{eq: Rotations}
    \begin{aligned}
        \psi_f = \left(\prod R_j \right) \psi_0 , \quad R_j= e^{-i\frac{1}{2} \beta_j (\hat{B}_j \cdot \vec{\sigma})}
    \end{aligned}
\end{equation}
whose result can be efficiently calculated by the Euler-Rodriguez formula.
Here $\beta_j = \gamma |B_j|\tau_j$ is the duration in angle of precession.  Constraining $\vec{B}$ to the $y$-$z$ plane, we 
write $\hat{B_j}$ using the angle $\theta_j$ relative to $\hat{z}$.  

\red{We note that this geometric approach should work for any system that can be treated in terms of Pauli rotations on the Bloch sphere.  For instance resonant pulses for 
dipolar or hyperfine transitions can be treated in this formalism by adjusting the strengths $B_\mathrm{RF}$ 
and detunings $\delta \omega$ of the RF pulses. Calculated in the rotating-wave 
approximation, the mapping is $\theta_j = \arctan(B_\mathrm{RF,\ j}/
(\delta\omega_j/\gamma))$. Manipulating $m_f$ hyperfine 
states is the same, but with the usual adjustment to the 
gyromagnetic ratio. In the resonant case, the pulse 
sequences we 
develop here provide robustness to errors in the detuning 
or the Larmor frequency, the strength of the RF field, and 
the duration of the pulse. Since this approach to pulse 
design is fundamentally geometric it should 
extend to other 2-state systems with overlapping qubits.}

The complete error-correcting sequences are difficult to picture, but an intuitive 
description can be built from single-segment two-species pulses, and error-correcting 
single-species sequences. 
We find a single segment pulse with $\theta_1 \approx 72.2^\circ$ that is only $7\%$ 
below the Mandelstam-Tamm quantum speed limit, and another with $\theta_1 \approx 
45.3^\circ$, which is insensitive to errors in pulse area, as demonstrated 
experimentally in Figure \ref{fig:45v70}.

\begin{figure}[h]
     \centering
\subfloat{
    \includegraphics[trim={1cm 0cm 0.2cm 1.5cm}, clip, width=0.65\linewidth]{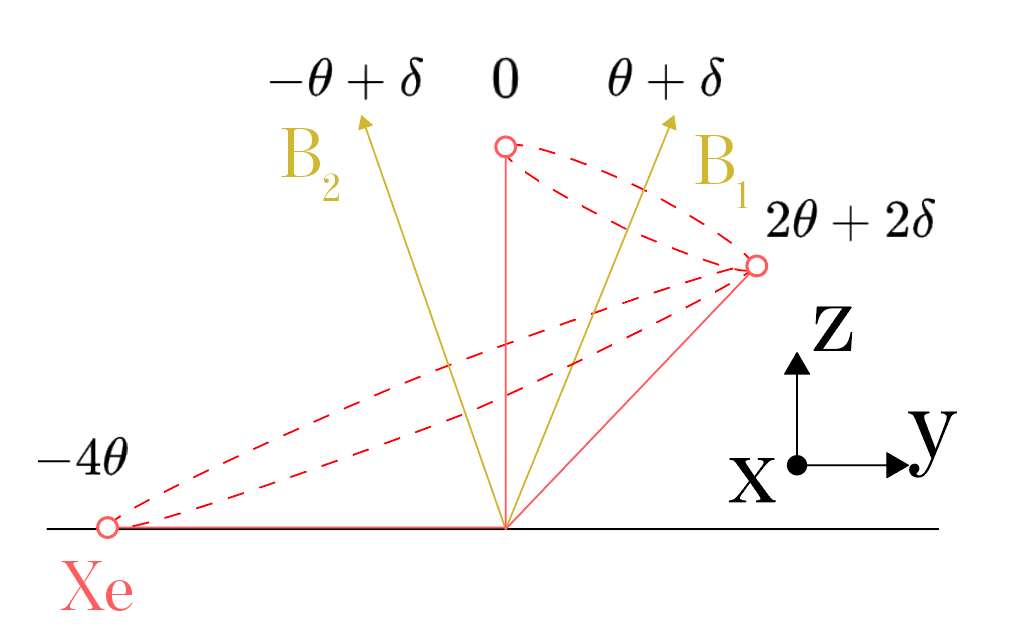}%
    } 
    \begin{minipage}{0.22\linewidth}
    \vspace{-3.3cm}
    \hfill
\subfloat{
         \centering         
    \includegraphics[trim={1.5cm 1cm 1.8cm 1cm}, clip, width=0.99\linewidth]{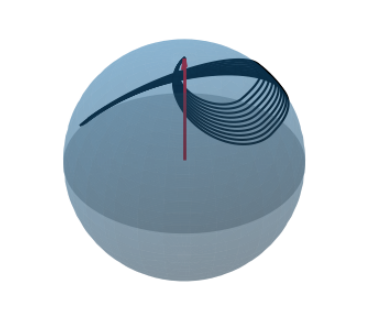}% 
    }
    \\
\vspace{-0.3cm}
    \subfloat{
         \centering         
    \includegraphics[trim={1cm 0.9cm 1.1cm 0.7cm}, clip, width=0.99\linewidth]{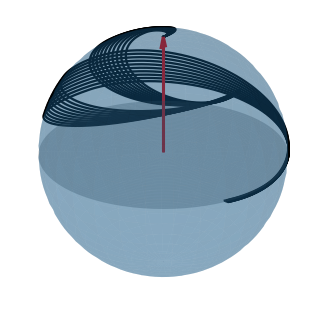}}
    \end{minipage}
    \caption{\emph{Left}: Diagram of a pulse robust to $B_y$. The pulse consists of $B_1$ and $B_2$ applied sequentially for equal durations; see table \ref{tab: pulse table}. A background field error shifts both $B_1$ and $B_2$ clockwise from 
    the intended angle $\theta$ by $\delta$.  The ensemble is shown at 
    three locations: the initial state, after the first 
    segment (2$\theta + 2\delta$) and the final state.  The error in $B_2$
    cancels the error in the spin state following pulse $B_1$.  \emph{Right}:  Bloch 
    sphere trajectories. The ensembles begin in the up-state and curves are computed for  
    background field errors of $\pm 20$nT. \emph{Above}: Two segment pulse canceling $B_y$ errors.  \emph{Below}: Four segment pulse canceling $B_y$ and $B_z$ errors. The cusp halfway through is when the $\hat{z}$ field reverses.}
    \label{fig:2+}
\end{figure}

Developing control pulses that are robust to background magnetic fields 
is crucial to our application. Sequences with this property have not seen extensive  
study, as they are only relevant at low fields. 
Background fields modify both the initial state and the 
direction and strength of the applied field, combining to change $
\theta$ at leading order.  We therefore compensate with composite pulses that 
cancel errors in $\theta$ between segments. 
By symmetrizing the pulses across $\hat{z}-\hat{x}$, we 
suppress errors due to fields along $\hat{y}$, and vice-versa. 
Figure \ref{fig:2+} illustrates the approach.  Two species pulses are found numerically starting from our single species 
solution, subject to those symmetry constraints.  The solutions for specific pulses are provided in Table \ref{tab: pulse table}.

\begin{center}
\begin{table}
\begin{tabular}{ l | c | c } 
  Robust to & $\theta$ (${}^\circ$) & $\ \beta\ $ (${}^\circ$)  \\ 
  \hline
  N/A: Fastest & 72.2 & 95.90  \\ 
  Pulse Area&  45.3 &  191.8   \\  
  $B_y$; Area  &  [22.8, -22.8] &  191.8  \\  
  $B_z$; Area &  [22.6, 180-22.6] &  191.8 \\ 
  $B_z$; $B_y$; Area &  \ [11.4, 191.4, 168.6, -11.4]\  &  191.8\\ 
\end{tabular}
\begin{minipage}[l]{1\linewidth}
        \includegraphics[width=0.9\linewidth]{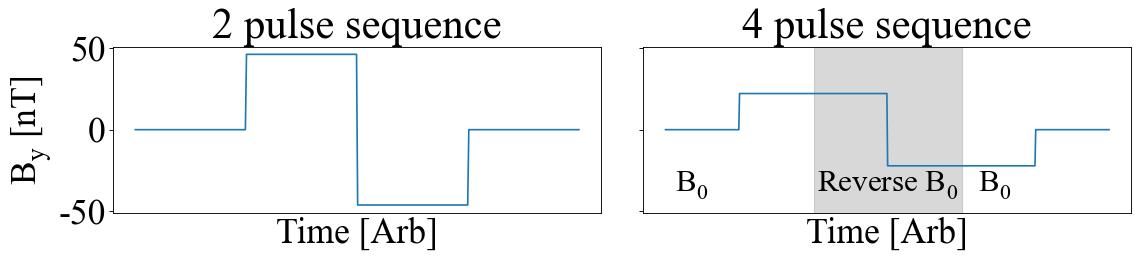}
        \label{fig:side_image}
    \end{minipage}
  \caption{Pulse sequences in terms of $\theta$ and $\beta$ for each segment $j$.  These are all pulses that simultaneously control both xenon and helium.  Time-domain of B$_y$ shown below, for both $B_y$ correcting pulse sequences, assuming B$_0 = 110$nT.}  \label{tab: pulse table}
  \end{table}
\end{center}

This protocol allows independent tuning of the final states for the two ensembles.
The two-species sequences either have all $\beta_n > \pi$, or 
else all $\beta_n < \pi$. Since the deviation from $\beta_n=\pi$ is 
nearly the same 
for the two spin ensembles, i.e. $3\pi - 
\beta_n(\mathrm{He}) \approx 
\beta_n(\mathrm{Xe}) - \pi $, they have nearly equal but 
opposite sensitivities 
to $B_x$, as shown in figure  \ref{fig:bangbang}.   
Adjusting $B_x$ tunes $\alpha$ for  helium and xenon in 
    opposite directions, and 
adjusting $\theta_n$ tunes them in the same direction.

\begin{figure}
\centering
\subfloat{
    \centering
    \includegraphics[trim={0cm 0cm 0 0}, clip, width=0.9\linewidth]{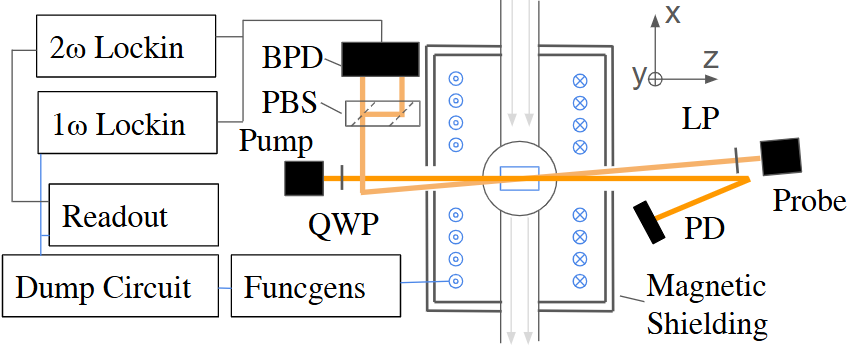}
    }\\
    \subfloat{
    \centering
    \includegraphics[trim={9.2cm 6.8cm 7.2cm 6.3cm},clip, width=0.9\linewidth]{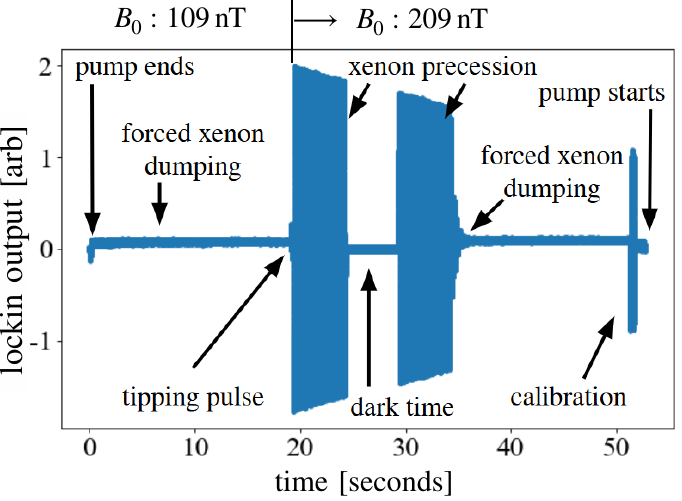}%
    }
\caption{\emph{Above}: Schematic of our apparatus. The cell contained {${}^{87}$Rb} metal and {${}^{129}$Xe}, {${}^{3}$He}, and N$_2$ 
gasses.  Lasers were on the D1 line. 
    The nuclei were polarized to between 8nT and 100nT as desired for the measurement.
   The cell was 2mm x 2mm and $B_0$ was around 110 nT before the tip, and 210 nT after the tip. The cell and shielding were made by Twinleaf. The laser power  was stabilized following the linear polarizers.  BPD - Balanced Photodiode, PBS -- Polarizing beam splitter, QWP -- quarter wave plater, LP -- linear polarizer, PD -- photodiode.
   \emph{Below}: Raw data from a single measurement. The y-axis is the in-phase output of the $1\omega$ lockin-amplifier.  During ``dumping'' feedback forces the nuclei to lie along the pump axis. The dark time allows rubidium-polarization effects to be studied.
   } \label{fig: exptPRL}
\end{figure}

\begin{figure}
    \subfloat{
    \includegraphics[trim={2.25cm 5.05cm 18.1cm 5.3cm}, clip, width=0.48\linewidth]{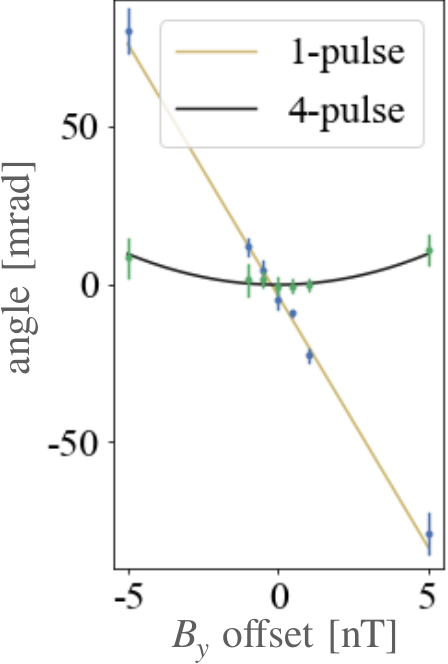}
    }
    \hfill
    \subfloat{
    \includegraphics[trim={0.75cm 0.75cm 0.75cm 0.83cm}, clip, width=0.45\linewidth]{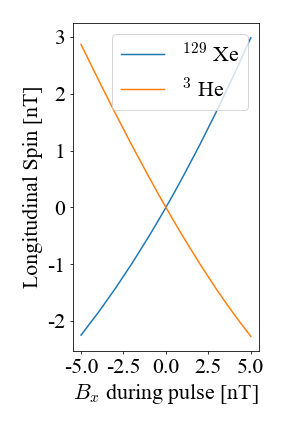}%
    } 
        \caption{\emph{Left}: Measured $\alpha$ for single and robust pulses as background fields are applied along $\hat{y}$.  The line is a single-parameter fit to equation \ref{eq: Rotations}, with only the polarization left free. The points are the average of 6 measurements with the error bar their standard deviation.  For the single pulse, $\alpha$ varied by $\sim 150$ mrad, while it varied by $\lesssim 10$ mrad with the robust pulse. \emph{Right}: Calculation of the effect of an error in $B_x$. The nearly equal and opposite behavior  allowed independent tuning of the ensembles.}
    \label{fig:bangbang}
\end{figure}

To validate the protocol experimentally and test whether it can reach $\alpha<10^{-3}$, we apply these pulses on co-located, 
{spin-exchange-optically-pumped} ${}^{129}\mathrm{Xe}$ and ${}^{3}\mathrm{He}$ ensembles using the 
apparatus sketched in figure \ref{fig: exptPRL}. 

The system features a 
novel read-out system capable of sensing both 
$\langle S_z \rangle$ 
and $ \langle S_y \rangle $ of the nuclei, a  
crucial feature for assessing our pulse sequences. 
 Typical read-outs isolate $\langle S_y 
\rangle$ and are only quadratically sensitive 
to small $\alpha$. To our knowledge 
 this is the first system capable of measuring 
 both components of the nuclei 
both before and after a 
 state-transition.

The nuclear spins ($\vec{\Xe}$) are measured via their spin exchange interactions with the rubidium vapor ($\vec{\Rb}$) as described by 
\begin{equation*}\label{eq: blocheqn}
\begin{aligned}
    &\frac{d }{dt} \left( Q(\Rb) \vec{\Rb} \right) =
     \gamma_{\Rb} \vec{B} \times \vec{\Rb} +  \Gamma \langle \phi \rangle \vec{\Xe} \times \vec{\Rb} \\
    & -\Gamma \langle \phi^2 \rangle \frac{\hbar}{2}\left( 
    \frac{\vec{\Rb}}{\Rb_\mathrm{max}} - \frac{\vec{\Xe}}{{\Xe_{\mathrm{max}}}} \right) 
   - \overleftrightarrow{D} \cdot \vec{\Rb} - R \left(\vec{\Rb} - \Rb_\mathrm{max} \hat{z} \right).
\end{aligned}
\end{equation*}
The equation of motion 
for $\vec{\Xe}$ can be found by swapping $\Xe \leftrightarrow \Rb $ and dropping the final term.  The first term is Larmor precession, the 
second term is first-order spin-exchange between alkali and nuclei, and the third 
term is second-order spin-exchange.  $R$ is the optical-pumping rate, and $\overleftrightarrow{D}$ are the decay rates of 
the alkali polarization.  $Q(K)$ is the slowing-down factor\cite{Appelt1998theory}.

First-order spin-exchange acts 
as an effective magnetic 
field aligned with the nuclear spins 
\begin{equation}\label{eq: nt}
b_S \langle \vec{S} \rangle  \equiv 4 \mu_0 
 \kappa_{\mathrm{Rb}S} \rho_s \langle \vec{S} \rangle \gamma_S/
 3. 
 \end{equation}
with $ \kappa_{\mathrm{Rb}S} = \frac{3 \Gamma_S \langle \phi_S \rangle }{ 4 \mu_0 \rho_S \gamma_\mathrm{Rb} \gamma_{S}}$ describing the strength of the interaction, where $\Gamma$ is the rate of collisions, $\phi$ the phase angle due to the collision and $\rho_S$ the nuclear spin density.
 Second-order spin-exchange polarizes the 
nuclear spins during the pumping period. 

We monitor $\Rb_z$ by Faraday rotation of an off-resonant 
probe beam, and extract $\vec\Xe$ by comparing with magnetic 
calibration data. 
Sensitivity to both $\hat{S_y}$ and $\hat{S_z}$ fields is obtained by applying 
a square-wave magnetic field modulation along $\hat{y}$ at $
\omega = ${1567 Hz} and demodulating the $
 \hat{z}$-component 
of the rubidium at both $1\omega$ and $2\omega$ with a 
pair of 
lock-in amplifiers.  In the absence of a field or spin along $\hat{y}$, the rubidium 
response should be  
at $2\omega$, as the perturbation is symmetric around the 
pumping 
axis.   Therefore, the 1$\omega$ signal is directly 
sensitive to $S_y$, while the 2$\omega$ signal is 
primarily sensitive to $S_z$.

 The separation of the nuclear spin components onto the harmonics of $\omega$ is 
 imperfect at the few percent level, because the rubidium response also depends on the 
 alignment between the pump beam and the net field including the nuclei.  This is 
 related to the heading errors seen in optical magnetometry.
 Testing our pulses at the part-per-thousand level, and across a wide range of pulse 
 parameters therefore required a joint fit procedure.

To map the measured rubidium signal to the 
$\hat{S_y}$ and $\hat{S_z}$ we 
applied static and rotating magnetic fields and built a set 
of model functions describing the responses of our 4 recorded output 
channels. These are the two 
quadrature components of each of the $1\omega$ and $2\omega$ lock-in amplifiers, which we label $\ell_i$.
We extract $S_z$ and $S_\mathrm{rot}$ by finding the values that 
minimize the $\chi^2$ between the eight model functions 
$g_i(B_z, B_\mathrm{rot})$ and the eight measured $\ell_i$, with 
$B_\mathrm{rot}$ extracted at both the xenon and helium precession 
frequencies.  

Optical pumping of the nuclei was done 
at 80\% rubidium polarization, while read-out was done at 10\% to reduce back-action, 
which rotates the xenon back toward the pump axis\cite{Kornack2002}.  
The measurement period is shown in Figure 
\ref{fig: exptPRL} and caption. We also monitored cell temperature, 
the  laser powers before and after the cell, and the voltages of the tipping pulse. 
The temperature was stabilized near 115$^\circ$ 
by a cascade PID system.  The lasers (following polarization) were 
stabilized by feedback to acousto-optic modulators.  We aligned the pulse coils and 
background field 
by minimizing the xenon precession frequency and aligned the lasers with the coils by 
minimizing xenon 
excitation when switching off the laser. 

We now turn to the key experimental question: can we prepare the target state with  
milliradian repeatability and tunability, sustained over multiple 
hours? This is the level required to 
fully leverage noble-gas $T_2$ times of 7000 seconds in next-generation tests of 
fundamental physics. Special efforts were required to demonstrate this level precision 
as not only must  the state-preparation be 99.9\% repeatable, but 
the read-out must be stable at that level as well. The following procedures were 
required.

\emph{1.} Sensitivity to $\langle S^2 \rangle$ was suppressed by measuring $\langle 
S_z \rangle$, since our target state is $\langle S_z \rangle=0$. This also provided linear 
sensitivity to $\alpha$.

\emph{2.} We used a 4-segment sequence robust to $B_y$, $B_z$ and 
pulse area. Data in Figure 
\ref{fig:bangbang} shows the error suppression works as predicted.  This is  
a parametric suppression, so it will be $\mathcal{O}
(10^5)$ for the $\sim$100 pT drifts seen in magnetically shielded 
environments. Comparable agreement with theory was seen for applied errors in $B_z$.

\emph{3.} We compared $B_z + b_S\langle S_z 
\rangle$ across a pump-transfer cycle to suppress slow drifts and subtract off $B_z$.  We measured   
after the pulse sequence and again following an extended 4-minute-long dark time. This 
was a compromise between a short baseline, and complete xenon decay.

\emph{4.} We partially linearized the read-out by shifting 
$B_z$ during state-transfer (see Fig. 3) such that the net 
longitudinal field 
($B_z + b_S\langle S_z \rangle$) experienced by the 
rubidium before 
and after the pulse matched the most linear part of our calibration to $10$nT.  This was crucial, as non-linear 
terms in the calibration were most sensitive to the 
environment. 

\emph{5.} We calibrated the read-out every 4 hours.  A measurement cycle took 6-8 
minutes giving us 
$\sim$30 measurements per calibration.  At twice the lifetime of the 
ultra-long-lived 
states, 4 hours is representative of the durations 
over which stability must be maintained. 

\emph{6.} We minimized ambiguity from helium by using  
pumping parameters that maximize xenon polarization and minimize helium 
polarization. Two species pulses were always applied. This allowed us to confirm the 
performance of our joint pulses and test for systematics, and also 
prevented the build-up of helium polarization.

\begin{figure}
     \centering
     \subfloat{
        \includegraphics[trim={0cm 0cm 0cm 0cm}, clip, width=0.9\linewidth]
        {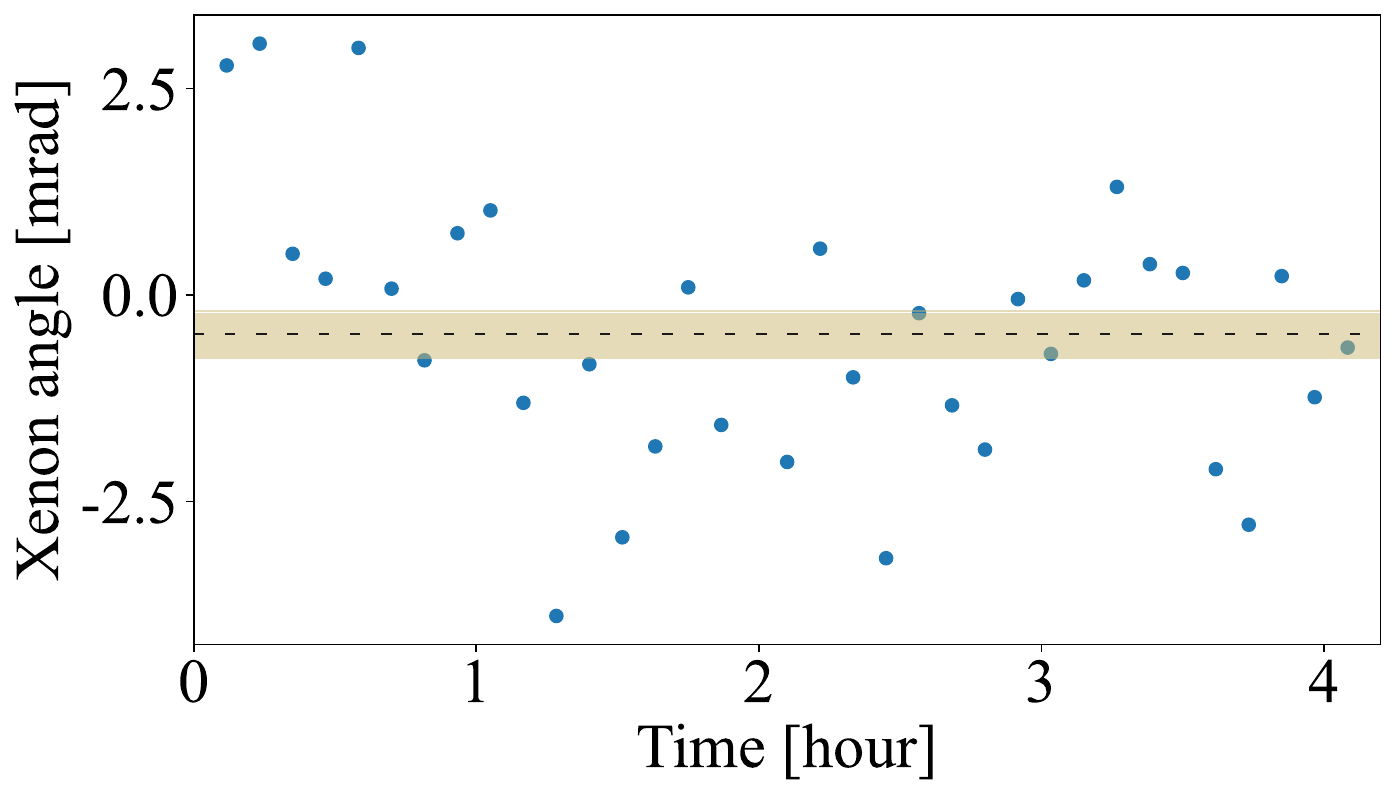}
        }
        \caption{Repeated measurements of xenon initial state 
        ($\alpha$) using \red{pulses robust against errors in $B_y$, $B_z$ and pulse area.  The changing $B_x$ tunes the pulse towards zero.   
        The yellow band shows the standard-error-of-the-mean, which was $0.5$ mrad across 30 
        measurements. This should be compared to the target angle of 0 mrad for which we tuned this pulse. The standard deviation of the measurements was consistently around 1.5 mrad, dominated by read-out error. The standard error of the mean reaches 0.5 mrad.}}
        \label{fig:stability}
\end{figure}

With these procedures in place we reliably achieved our target millirad precision 
over several hours and could tune our 
ensembles to a desired $\langle S_z \rangle$, as shown in 
figure \ref{fig:stability}.  The standard deviation of a multi-hour measurement was typically 1.5 mrad, and after averaging down we could reliably obtain 0.5 mrad.  The scatter and long-term drift of the data were both dominated by fluctuations in the read-out calibration, as determined by measuring the longitudinal magnetization of a single tip over time.  This further motivates a new approach to the readout system. This data demonstrates that our protocol can tune to and 
repeatably generate a target superposition at the level needed for the next 
generation of fundamental physics applications.  The exact tuning  
would be that which maximizes the coherent integration time in the specific 
apparatus. 

We also investigated measuring   
$\alpha$ absolutely.  This was a subtle exercise.  On top of the 
issues that made milliradian  
precision challenging, we found a systematic $\sim2$nT offset between the 
data after the tip 
and after the long decay time.  Ultimately, we traced this back to 
second-order spin-exchange between the xenon and rubidium, i.e. the $\langle \phi^2 
\rangle$ term in the rubidium equation of motion. 
This is the term that is used to pump the xenon via the rubidium.  Here, the 
reverse effect was in action, 
with the polarized xenon slightly pumping the rubidium during detection, 
and modifying the rubidium dynamics as compared to the calibration data.  

This appears to be a previously unreported effect of second-order spin-exchange. 
Scaling considerations point to nT level effects from the $\langle \phi^2 \rangle$ 
terms.  For $20 \%$ polarized xenon, we estimate a $50$Hz effect, while 
Larmor precession and optical pumping were around $3000$Hz.  The effect is therefore 
comparable to the Berry's phase and heading error systematics identified in another 
recent vector magnetometer that did not contain co-located nuclear 
spins\cite{Wang2025pulsed}.

\begin{figure}
     \centering
     \subfloat{
        \includegraphics[trim={0cm 0cm 0cm 0cm}, clip, width=0.9\linewidth]
        {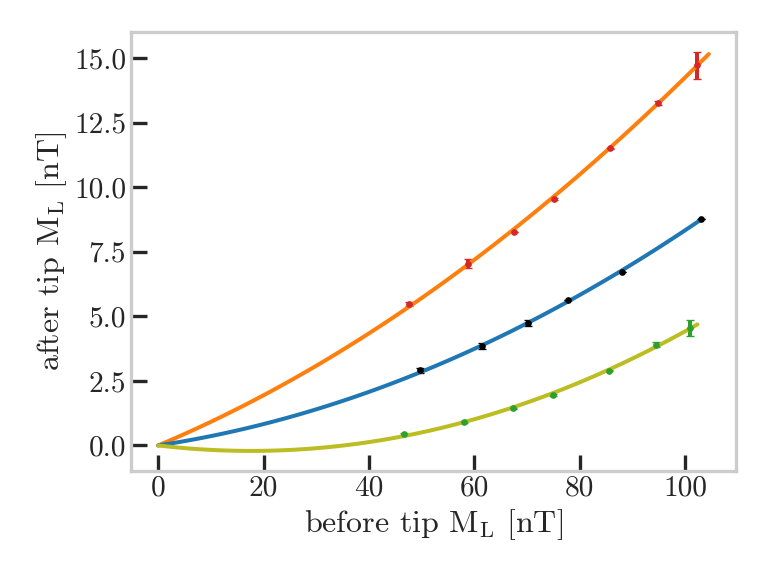}
        }
        \caption{Data scanning over nuclear spin polarization at three different tip angles that spanned $\sim$100 mrad around an optimal $\pi/2$ pulse (upper orange curve: smallest tip angle; lower green curve: largest tip angle).  The results match expectation to within experimental  uncertainty. The component proportional to $\langle S^2\rangle$, which produces the systematic shift, was consistent across tip angle. The component proportional component proportional to $\langle S_z\rangle$ before the tip, which represents the  tip-error, varied linearly with small changes in tip, as expected.}
        \label{fig: overtip undertip}
\end{figure}

The origin of the systematic was confirmed by looking for a characteristic $\langle 
S^2\rangle$ scaling of the systematic offset. This scaling cannot be due  
to a pulse error, which scales as $\langle 
S_z \rangle$ before the tip.  We measured the scaling by varying the  
nuclear-spin polarizations between 5\% and 20\%, as shown in figure \ref{fig: overtip undertip}.  The data were in agreement with full density matrix 
simulations, as long as  
back-polarization  
and incomplete cell pumping were included.  These terms are often neglected when 
analyzing coupled spin ensembles due to their small size\cite{Kornack2002}, however 
they must be accounted for at the sub-percent level. 

While in principle the back-pumping of rubidium by xenon could be modeled, in practice 
its precise value depends on  
the second-order spin-exchange parameters, which themselves 
depend on the environment and are 
difficult to measure precisely\cite{rice2002rubidium}.  Furthermore, the size of the 
systematic depends strongly on the lock-in phase and filter parameters.  
Taken together \red{with the desire for improved read-out stability} it seems more profitable to pursue other avenues 
for future improvements. 

In particular, either pumping the rubidium to saturation or leaving it unpumped would 
greatly reduce the systematic. While back-action from highly-polarizing the rubidium 
can be controlled by reversing the pump beam 
polarization, generically this cancels sensitivity to $b_S \langle S_z \rangle$ as 
well.  If the rubidium is left unpolarized, the 
 signal comes only from second-order spin exchange and there 
 is no background from $B_z$.  Instead, the challenge would be to 
 detect the small component of $\langle S_z \rangle$ on top of a large $\langle S_y 
 \rangle$, which rotates out of the $x-y$ plane in the presence of rubidium due to the 
 coupling of rubidium and xenon ensembles. 

In summary, this first attempt at error-corrected, 
simultaneous control of 
ultra-long-lived states achieved  
millirad state-preparation across several hours.  This is 
20-30 times better than 
previous efforts using resonant pulses, sharp pulses, or 
sudden B$_0$ shifts\cite{Terrano2019b}. 
This level of control will enable  
coherent phase integration over the lifetime of a  
noble-gas nuclear dipole oscillation at state-of-the-art 
experimental levels, and make 
possible the most sensitive measurements 
of energy splittings to date.  We also demonstrated, for 
the first time, protection of state-preparation errors  against spurious magnetic fields 
along all axes. 

In a broader context our results are comparable with the first attempts 
at high-fidelity quantum control, which began a decade ago in  
hyperfine states\cite{Ballance2016high} and have since improved by another 1000-
fold\cite{An2022, Smith2025single}.  
Looking forward, the required level of state-preparation will become 
more stringent as measurements of the phase of the ensemble improve with improvements 
in noise, read-out and ensemble 
size.  Further progress on preparing these nuclear-dipole states will therefore be 
needed.  Ultimately, a 
spin-projection-noise-limited measurement with these ultra-long-lived nuclear spin 
states --  plausibly the most sensitive 
possible measurement of quantum mechanical energy splittings -- will require control 
protocols capable of state-preparation down to the 10 nanoradian level.  

\noindent
{\bf Acknowledgements} \newline \indent
We thank M.E. Limes for discussions on optical magnetometry and C. Arenz for 
discussions of dynamical stabilization and simultaneous control. We thank Jacob Feltman for feedback on the 
manuscript. This work was supported by ASU, DOE Grant DE-SC0025564 and the ASU Quantum 
Initiative. 

\section{Data Availability}

The data that support the findings of this article are openly available \cite{data}.

\appendix 

\setcounter{section}{0}
\setcounter{equation}{0}
\setcounter{figure}{0}
\setcounter{table}{0}

\renewcommand{\thesection}{S\Roman{section}}
\renewcommand{\theequation}{S\arabic{equation}}
\renewcommand{\thefigure}{S\arabic{figure}}
\renewcommand{\thetable}{S\arabic{table}}

\section{Supplemental: Extracting $M_\mathrm{T}$ and $M_\mathrm{L}$ from a single probe beam}

In order to measure the components $ \langle S_\mathrm{rot} \rangle$, $
\langle S_z \rangle$ of the xenon spin state, we apply a square-wave 
magnetic field modulation at $\omega = 1567$Hz and measure the 
response of the rubidium $ \langle K_z \rangle$. This response is 
demodulated with respect to $1 \omega$ 
and $2 \omega$ using SRS 865A lock-in amplifiers, providing 4 total lock-in outputs which we label $\ell1x,\, \ell1y,\, \ell2x,\, \mathrm{and}\ \ell2y$. 

In the presence of a single precessing nuclear spin, the response of 
each lock-in output can be decomposed into parts: A DC shift and an 
oscillation at the nuclear spin frequency. For two nuclei, therefore, 
each measurement will involve three data points per lock-in 
output for a total of 12 data points.

Concretely, these are: 
\begin{enumerate}[noitemsep, leftmargin=1.3cm]
\renewcommand{\labelenumi}{(\the\numexpr 2*\value{enumi}-1 \relax, \the\numexpr 2*\value{enumi} \relax)}
    \item the $x$ and $y$ offsets of the $1\omega$ 
lock-in
    \item the amplitudes at xenon frequency for the $x$ and $y$ components of the $1\omega$ lock-in
    \item the amplitudes at helium frequency for the $x$ and $y$ components of the $1\omega$ lock-in
    \item the $x$ and $y$ offsets of the $2\omega$ lock-in
    \item the amplitudes at xenon frequency for the $x$ and $y$ components of the $2\omega$ lock-in 
    \item the amplitudes at helium frequency for the $x$ and $y$ components of the $2\omega$ lock-in 
\end{enumerate}

To calibrate the responses, we applied static fields along $\hat{z}$ and rotating fields in the $
\hat{x} - \hat{y}$ plane at 7\,Hz.
The static field was set at 12 
different values between $-315\,$nT and $-95\,$nT,  and the rotating field set to 7 different 
amplitudes  between 0 and 150\,nT.  The calibration sequence 
was repeated 2-6 times, and outliers dropped. 

For each of the 84 combinations of static and 
rotating fields, we analyzed the $\ell1x,\, \ell1y,\, \ell2x,\, \mathrm{and}\ \ell2y$ signals for 
their static (DC) and rotating (7Hz) components.  We then 
construct a set of eight model functions $g_i$ that give the expected DC or rotating 
response of each channel to a given applied 
field.  For example $g_{\ell1x; \mathrm{DC}}(B_z, B_\mathrm{rot})$ would give the DC 
signal seen on channel $\ell1x$ given 
applied fields of $B_z$ and $B_\mathrm{rot}$.  The  model 
functions are polynomials in $B_z$ and $B_\mathrm{rot}$, 
including cross-terms, e.g.

\begin{equation} \label{eq: signals}
    \begin{aligned}
        g_{\ell i;DC} = & ( C_{00} + C_{10}M_L + \dots)\\
        & + (C_{02} + C_{12}M_L + \dots )M_T^2 \\
        & + (C_{04} + C_{14}M_L + \dots)M_T^4 + \dots \\
        g_{\ell i; 1\omega} = & (C_{01} + C_{11}M_L + \dots)M_T \\
        & + (C_{03} + C_{13}M_L + \dots)M_T^3 + \dots
    \end{aligned}
\end{equation}
which we find 
provide a good description of the calibration data.  The cutoff order for these polynomials was set to  the smallest such that data looks, by an educated assessment, to be 
properly modeled.   We settled on modeling the dc offset with 4th order in $M_T$ and 4th order in $M_L$, and the 1$\omega$ amplitude up to 5th order in $M_T$ and 2nd order in $M_L$. 

\begin{figure*}
    \centering
    \includegraphics[width=0.9\linewidth]{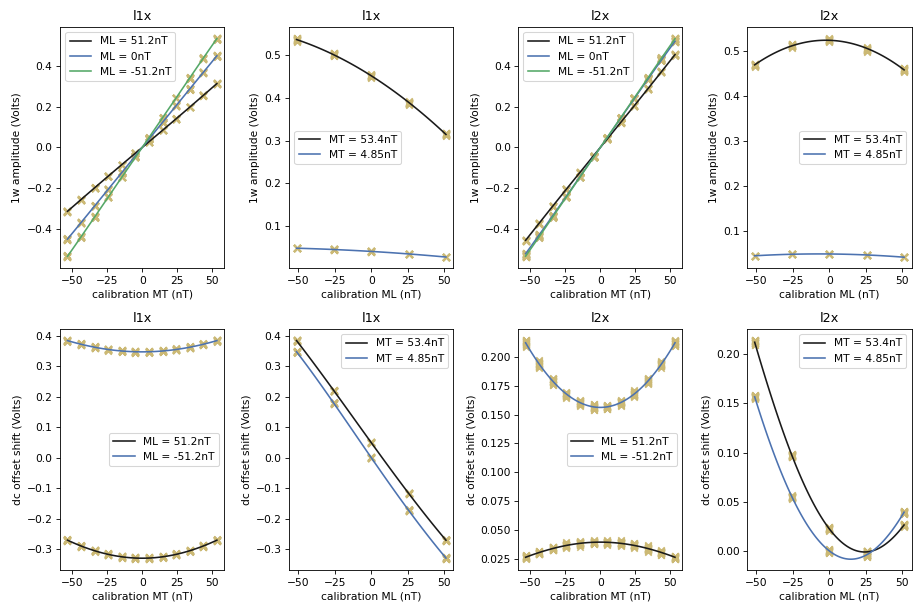}
    \caption{Fits of the DC and 1$\omega$ (7Hz) components for 2 of the 4 lockin signals to 
    their calibration strengths $M_T$ and $M_L$ in the rubidium-only cell. Each point has 
    been measured 25 times throughout the entirety of the calibration test. l1x stands 
    for the x output of the lockin for the first modulation harmonic, and l2x stands for 
    the x output of the lockin for the second modulation harmonic. Note how $M_T$ has the 
    strongest effect on the 1$\omega$ amplitudes, meanwhile $M_L$ has the strongest effect on 
    the dc offsets. Note also how the dependencies of l1x and l2x on $M_T$ and $M_L$ are 
    different in their shapes.}
    \label{fig:cal_fits}
\end{figure*}

Figure \ref{fig:cal_fits} shows an example of these polynomial 
fits to our calibration data for the rubidium only cell.  $M_\mathrm{rot}$ and $M_z$ are 
determined by matching the 8 or 12 lock-in measurements to the 8 calibration curves, with 
$M_\mathrm{rot}$ and $M_z$ as the free parameters in the equations of \ref{eq: signals}.  
In data with xenon only, there are 8 lock-in measurements, with xenon and helium there 
are 12. 

\newpage

\bibliography{references.bib}
\end{document}